\documentclass[prb,a4paper,superscriptaddress,floatfix,citeautoscript,preprint]{revtex4}
\usepackage{psfig,epsf}
\usepackage{graphicx}
\usepackage{rotating}
\usepackage{bm}
\usepackage{jpc}

\newcommand{\jacs}{J. Am. Chem. Soc.}
\newcommand{\pnas}{Proc. Natl. Acad. Sci. U.S.A.}
\newcommand{\jpca}{J. Phys. Chem. A}
\newcommand{\jpcb}{J. Phys. Chem. B}

\begin{document}

\title{A TDDFT study of the excited states of DNA bases and their assemblies} 

\author{Daniele Varsano}
\affiliation{Departamento de F\'{\i}sica de Materiales, Facultad de 
Qu\'{\i}micas, Universidad del Pa\'{\i}s Vasco and Donostia International 
Physics Center (DIPC),20080 San Sebastian (Spain)}
\affiliation{European Theoretical Spectroscopy Facility (ETSF)}
\author{Rosa Di Felice}
\affiliation{National Center on nanoStructures and 
Biosystems at Surfaces (S3) of INFM-CNR, Via Campi 213/A, 41100 Modena, Italy}
\author{Miguel A.\,L. Marques}
\affiliation{Centro de F\'\i sica Computacional, Departamento de F\'\i sica da
Universidade de Coimbra, Rua Larga, 3004-516 Coimbra, Portugal}
\affiliation{European Theoretical Spectroscopy Facility (ETSF)}   
\author{Angel Rubio}
\affiliation{Institut f\"ur Theoretische Physik, Freie Universit\"at Berlin,
Arnimallee 14, D-14195 Berlin, Germany}
\affiliation{Departamento de F\'{\i}sica de Materiales, Facultad de 
Qu\'{\i}micas, Universidad del Pa\'{\i}s Vasco and Donostia International 
Physics Center (DIPC),20080 San Sebastian (Spain)}
\affiliation{European Theoretical Spectroscopy Facility (ETSF)}

\begin{abstract}
  We present a detailed study of the optical absorption spectra of DNA bases and
  base pairs, carried out by means of time dependent density functional theory.
  The spectra for the isolated bases are compared to available theoretical and
  experimental data and used to assess the accuracy of the method and the
  quality of the exchange-correlation functional: Our approach turns out to be
  a reliable tool to describe the response of the nucleobases. Furthermore, we
  analyze in detail the impact of hydrogen bonding and $\pi$-stacking in the
  calculated spectra for both Watson-Crick base pairs and Watson-Crick
  stacked assemblies. We show that the reduction of the UV absorption intensity
  (hypochromicity) for light polarized along the base-pair plane depends
  strongly on the type of interaction. 
  For light polarized perpendicular to the basal plane, the hypochromicity effect is reduced, 
  but another characteristic is found, namely a blue shift of the 
  optical spectrum of the base-assembly compared to that of the isolated bases. 
 The use of optical tools as fingerprints for the
  characterization of the structure (and type of interaction) is extensively
  discussed.
\end{abstract}

\maketitle

\section{Introduction}

Optical absorption and circular dichroism are widely used experimental tools to
characterize the structural and dynamical properties of biomolecular systems. 
The particular merit of optical characterization tools is that they can discriminate 
between intrinsic molecular properties and solvent-induced effects. 
This is especially important for DNA and
DNA-based compounds \cite{ruzsicska,crespo,daniels}. 
Traditionally, DNA molecules have always retained a special place in scientific investigation, 
for biological/medical issues. Lately, DNA is also attracting interest for several potential 
applications in the field of
nanotechnology, due to its stability (in solution), to its one-dimensional
character, and to the regular $\pi$-stacking, along with the unique properties of
self-assembling and recognition 
\cite{porath,endres,diventra_review,hezy,difelice02}.  
In relation to nanotechnology exploitation, the determination and 
interpretation of the electronic properties of nucleobases and of DNA helical arrangements 
are an extremely valuable foreword, and notable multidisciplinary efforts are currently 
devoted to such goals: we refer the reader to recent reviews about electronic 
structure calculations and possible charge motion behaviors \cite{porath,rosa_elsevier}. 
Furthermore, the knowledge of the
electronic properties, excited state lifetimes, and ultra-violet (UV) absorption
spectra is of paramount importance for our understanding of, e.g., the crucial
phenomenon of UV radiation-induced DNA damage \cite{kraemer,odom,crespo_nature}. From this
brief preamble, it is clear that the full characterization of the optical properties
of DNA molecules and DNA-based complexes is of great interest. 
In order to relate the optical
properties of the nucleic acids to their structure, spatial conformation, and
type of intra-molecular interactions, a valuable preliminary step is to gain insight into 
the excited-state properties of their building blocks, namely the
monomeric bases, and to understand the role 
of hydrogen-bonding and stacking when these monomeric units 
form complex assemblies. In their natural environment, the DNA bases are paired via
hydrogen-bonds in the Watson-Crick scheme \cite{watson},
and are covalently bonded to the sugar-phosphate backbone. The 
hydrogen-bonded base pairs interact with 
each other in the typical helical arrangement by 
inter-plane van der Waals forces \cite{hall}. 
To disentangle how the different interactions control
the DNA dynamics upon light absorption, it is important to infer how the
spectrum of a given isolated nucleobase is modified by mutual interactions in the
different spatial conformations of DNA-assemblies. This is the goal of the
present work: To provide a systematic study of the stacking and H-bonding
interaction effects in the optical spectra of molecular complexes formed with
isolated nucleobases. We undertake this task by means of
time-dependent-density functional theory (TDDFT) 
calculations \cite{runge_gross,gross_dobson,fiolhais,onida02}. 
 
TDDFT is gaining increasing popularity as an efficient tool to calculate
electronic excitations in finite systems, thanks to its simplicity and moderate
computational cost. Since the very beginning, TDDFT has provided good
results for the optical response of a large set of molecular
systems \cite{furche}, including some biomolecules (e.g., green fluorescent
protein chromophores and their mutants \cite{marques03,marques05}, 
chlorophylls \cite{sundholm00,sundholm03}, 
flavins \cite{neiss} and nucleic acid bases \cite{kaxiras,shukla04}, among many
others). It is still perhaps premature to discuss the general level of accuracy
of TDDFT when applied to biomolecules, especially when dealing with van der
Waals complexes \cite{rothlisberger}. Furthermore, as the field is quite recent,
we can still expect rapid methodological developments, in particular toward the
derivation of better exchange-correlation functionals. However, the existing
results are already promising, and pave the way for a broader application of TDDFT
in biochemistry.

Before studying complex helical DNA-based biopolymers, extensive tests are 
required to prove the performance and predictive power of TDDFT for 
DNA-based systems. One
of the aims of the present work is to provide results from first-principles calculations of
simple DNA-based assemblies that can be used as reference for future developments
and studies. This reference set includes the isolated DNA bases, hydrogen bonded
Watson-Crick pairs, a stacked guanine-cytosine dimer, and a quartet formed by 
two stacked guanine-cytosine pairs.

For the isolated DNA bases there are plenty of experimental and theoretical
results concerning their optical response\cite{crespo}. We do not carry out a
full systematic analysis of all the published data for the isolated DNA-bases,
but only those more pertaining to our investigation. The 
computational tools applied to nucleobases range from single excitation
configuration interaction (CIS), to complete active space 2nd-order perturbation
theory (CASPT2), and include other approaches such as multireference
perturbation configuration interaction (CIPSI).  A review of the performance of
such methods on the nucleobases \cite{crespo}, as well as relevant experimental
data can be found in Refs.~\onlinecite{fulscher97,fulscher95,lorentzon}.

Despite the large number of works on the excited states of isolated DNA bases, 
a limited number of studies is devoted to base pairs and base assemblies due to their
complexity. Shukla and Leszczynski studied adenine-uracil \cite{shukla02b},
adenine-thymine (AT) and guanine-cytosine (GC) pairs in the Watson-Crick configuration
\cite{shukla02} at the CIS level; Sobolewski and Domcke \cite{sobolewski04}
studied the low lying energy part of the spectrum of GC base pairs with the more
sophisticated CASPT2 technique; Wesolowski \cite{wesolowski} used an
embedding method to study the lowest excited state of the GC and AT pairs.  Very
recently, Tsolakidis and Kaxiras \cite{kaxiras} computed the whole absorption
spectra of the GC and AT pairs in different tautomeric forms of the 
bases in the TDDFT framework.

Moreover, very few studies exist on the excited-state properties of bases in a
stacked configuration. Jean and Hall studied the fluorescent properties of dimers
of 2-aminopurine stacked with DNA bases \cite{jean01} in different forms and
stacked trimers containing 2-aminopurine \cite{jean02}, showing the relevance
of the stacked geometries in the character of
the excited state transitions. To the best of our knowledge, the present results
are the first ab-initio calculations dealing with stacked natural pairs. 
Yet, we remark that the present results are for free standing nucleobase complexes,
i.e., not including solvation effects. These effects are known to be more
important for $n\pi^*$ than $\pi\pi^*$ type
transitions \cite{mennucci,mishra,shukla00,fulscher97,shukla02,shukla02c}.

The paper is organized as follows: In Sect.~\ref{method} we provide the details
of the real-time real-space method\cite{octopus} used to compute the optical
properties.  In Sect.~\ref{iso_bases} we present our results for the
isolated gas-phase nucleobases, i.e., guanine, cytosine, adenine, thymine, and
uracil, (G, C, A, T, and U, respectively). In order to understand the role of
hydrogen-bonding in shifting and modifying the spectral features, we show, in
Sect.~\ref{wats_crick}, results for H-bonded Watson-Crick GC and AT base
pairs (labeled GC$_{\rm H}$ and AT$_{\rm H}$, respectively). In
Sect.~\ref{stacking} we present results for a GC stacked dimer (labeled
GC$_{\rm S}$) that mimics the arrangement between C and G in two consecutive
planes in the real DNA double helix. We also discuss the relative roles of
$\pi$-stacking and hydrogen bonding in the optical absorption of a stacked
quartet made of two adjacent Watson-Crick GC pairs as in A-DNA [labeled d(GC)],
that combines hydrogen-bonding and $\pi$-stacking. In Sect.~\ref{conclusion} we
summarize the results of the present work and provide some perspectives for
future studies.

\section{Computational Framework}
\label{method} 

\subsection{Optical spectra by TDDFT}

The optical absorption spectra of DNA bases and base pairs were computed within
a real-space real-time version of TDDFT, as implemented in the code {\tt
  octopus} \cite{octopus}. This method does not rely on perturbation theory and
is competitive with implementations in the frequency
domain\cite{oct-rev,yabana96,yabana99,yabana99b}. The theoretical background and the computational
details of our scheme are extensively described elsewhere
\cite{octopus,oct-rev,rubio96,rubio97}: here we just summarize the crucial aspects
relevant to the present study of DNA complexes.

The starting point is the calculation of the ground state electronic structure
in the DFT framework, which is done within a pseudopotential approach (see below
for details). In order to access the excited state properties, the ground state
is then instantaneously perturbed with an electric field of magnitude $k_0$
along the three principal Cartesian directions (i.e., by applying the external
potential $v({\bm r},t)=-k_0 x_{\nu} \delta(t)$, where $x_{\nu}=x,y,z$).  The
amplitude $k_0$ must be sufficiently small in order to keep the system dipolar
response linear.  In this impulsive approach, all the frequencies of the system
are excited with the same weight.  Next, the time dependent Kohn-Sham (KS)
orbitals are propagated for a finite time, and the dynamical polarizability is
obtained as:
\begin{equation}
  \alpha_{\nu}(\omega)=-\frac{1}{k_0}\int\!\!d^{3}r\: x_{\nu} \delta n({\bm r},\omega)\,,
  \label{alpha}
\end{equation}  
where $\delta n({\bm r},\omega)$ is the Fourier transform of the 
relative density $n({\bm
  r},t)-n_0({\bm r})$, and $n_0({\bm r})$ stands for the ground state density.
The photo-absorption cross-section, the quantity usually measured in
experiments, is proportional to the imaginary part of the polarizability
averaged over the three space directions:
\begin{equation}
  \sigma(\omega)=\frac{4\pi\omega}{c} \Im \frac{1}{3} \sum_\nu \alpha_{\nu}(\omega),
  \label{sigma}
\end{equation}
where $c$ is the velocity of light.
Another widely used quantity is the dipolar
strength function $S_\nu(\omega)$\cite{onida02,octopus}, which is connected to
$\alpha_\nu(\omega)$ by
\begin{equation}
  S_\nu(\omega) = \frac{2m}{\pi\hbar^2}\Im\alpha_\nu(\omega)
  \,.
\end{equation}
With this definition, the Thomas-Reiche-Kuhn sum rule integrates to the number
of electrons in the system. $S_\nu(\omega)$ is the quantity plotted in our figures.

A significant advantage of this real-time approach is the fact that only
occupied states are needed, thus avoiding the calculation of the unoccupied
states that instead enter the traditional orbital (occupied-empty)
representation of the linear response equation\cite{casida}. The complete set of
empty orbitals required in the latter approach is fully accounted for by the
time propagation.  In addition, it is important to note that in the time domain
only the approximation to the exchange-correlation potential $v_{\rm xc}$ is
required, whereas in conventional frequency-domain linear response TDDFT
formulations also the $f_{\rm xc}$ kernel is needed \cite{runge_gross,gross_dobson,fiolhais,onida02,yabana96,yabana99,yabana99b}.
 For $v_{\rm xc}$ we employed the local density approximation (LDA) with the
Perdew-Zunger parametrization \cite{perdew81}. In the case of the isolated guanine
base, we also performed the time propagation with the generalized gradient
approximation (GGA) PW91 \cite{perdew96} $v_{\rm xc}$ functional.  As expected
\cite{marques01}, no substantial differences were detected between the two
parameterizations. Therefore, and as the LDA functional is numerically the most
stable, we chose to use this functional in the time propagation for all the other 
investigated systems \cite{note_functionals}.

Concerning technical details: We used a real-space grid made of overlapping
spheres with a radius of 4\,{\AA} centered around each nucleus. All quantities
were discretized in a uniform grid with spacing 0.23\,{\AA}. The time-step for
the time evolution was 0.0066\,fs (which ensures stability in the time-dependent
propagation of the KS wave functions) and the total propagation time was at
least 20\,fs. Such simulation parameters ensure well converged absorption
spectra up to about 8\,eV. The energy resolution, dictated by the total
simulation length by $\Delta E \sim h/T$, is better than 30meV. This sets the 
lower limit for the linewidth of the calculated spectra.
The electron-ion interaction (both in the
time-dependent and time-independent DFT calculations) was modeled by
norm-conserving pseudopotentials \cite{troullier}.

\begin{figure}
  \begin{center}
    \includegraphics[width=7cm]{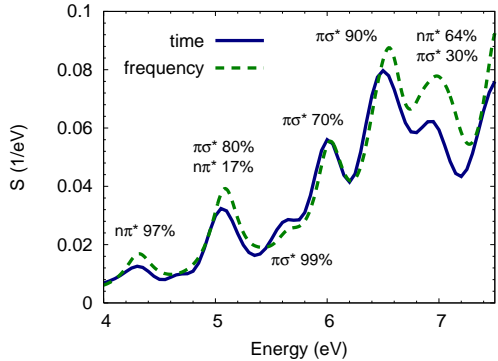}
  \end{center}
  \caption[]{Dipole strength function of 
    cytosine along the axis perpendicular to the base plane, calculated through
    the real-time propagation (blue solid curve) and in the frequency domain
    \cite{casida} (green dashed curve). An artificial broadening of 0.15eV
was adopted as an eye help to compare time and frequency domain spectra.
    Note that we can distinguish the $n\pi^*$
    character in an indisputable way only for the lowest energy peak.  At higher
    energy the $n\pi^*$ character is less pure, due to the mixing of a large
    number of transitions. This fact is illustrated here for cytosine, but we
    observed it for all the investigated nucleobases.}
  \label{fig:casida}
\end{figure}

To compare with experiments where the molecules are not aligned with respect to
the applied field, we need to average the response along the three Cartesian
axes [cf. Eq.~(\ref{sigma})]. This result can be directly compared to other
available data and used to discriminate among different nucleobase
conformations. However, the analysis can be pushed even further: in the case of
planar molecules (isolated bases, GC$_{\rm H}$, AT$_{\rm H}$) one can easily
distinguish between in-plane and out-of-plane transitions. While the former are
prevalently $\pi\pi^{*}$ and have large oscillator strengths, the
latter are much weaker (by one or two orders of magnitude) and are usually
hidden under the stronger signals.  In order to unveil the character of the weak
transitions, the time-domain analysis is not sufficient: Thus, we also performed
linear response calculations in the frequency domain \cite{casida} for all
planar structures (isolated bases, GC$_{\rm H}$, and AT$_{\rm H}$ pairs).  In
this formalism, several pairs of occupied/unoccupied orbitals participate in
each excitation with different weights: it is then possible to qualitatively
assign to each excitation the predominant character of the contributing
transitions.  In Fig.~\ref{fig:casida} we show an example of the transition
character of spectral features, derived from our frequency analysis of
cytosine.  The dipole strength function of cytosine along the direction
perpendicular to the base plane, obtained by both the time-domain and
frequency-domain techniques, is plotted against energy. The agreement
encountered between the two techniques turns out to be good in
the accessible energy range. For all other planar systems, we present the
analysis of the transition character obtained in a similar way (namely, by comparing 
time-domain and frequency-domain features), but without
explicitly showing the figures for the spectra obtained by frequency-domain
TDDFT.

We note that in the frequency-domain the transitions are delta functions. 
In order to make a fair comparison with the time-domain propagation, we have 
artificially broadened both spectra in Figure 1 using a lorentzian function of 0.15
eV width. Note that for the rest of the spectra shown in the paper obtained
with the time-propagation scheme, we did not add any additional broadening and the 
resolution of 30meV is fixed by the duration of the time evolution. 
Any additional broadening comes from the coupling of the excited 
state to other excitations (e.g. Landau damping).

\subsection{Ground-state equilibrium configuration by DFT}

\subsubsection{Starting atomic configurations}

The starting atomic coordinates of the isolated DNA bases and of the AT$_{\rm
  H}$ pair were derived from standard structural parameters for DNA
\cite{arnott72}, whereas for the systems GC$_{\rm H}$, GC$_{\rm s}$, and d(GC) we
extracted the simulated fragments from a high-resolution X-ray crystal structure
of A-DNA \cite{gao}, after a classical force-field structural adjustment.  The
crystal is formed by short helices of double stranded DNA with the sequence
d(AGGGGCCCCT), a model for poly(dG)-poly(dC) DNA.  The H coordinates (absent in
the X-ray PDB file) were assigned using the HBUILD command \cite{brunger} of the
CHARMM package \cite{charmm1,charmm2}.  The CHARMM force field was also used for a raw
relaxation of the entire d(AGGGGCCCCT) structure. From this grossly relaxed
polymer we extracted the H-bonded and stacked GC fragments, that have been
subsequently relaxed by quantum simulations. 
The simulations allow us to have a complete set of base configurations that 
could be used to perform configurational sampling and extract structural 
broadening effects, the analysis of this issue is left for future work.

All our simulations were done for systems in the gas phase: 
Note that the effect of a solvent may be crucial when comparing the
computational results with experimental data \cite{mennucci,mishra}.
Furthermore, the sugar-phosphate backbone was neglected. This restriction should
not be relevant in the energy range we are interested in (3--7.5\,eV),
as the sugar and the phosphate contribution to the absorption spectra
only starts to be important above 7.0\,eV.

\subsubsection{Technical details of the DFT runs}

The DFT structural optimization was carried out using the B3LYP/6-31++G(d,p)
exchange-correlation functional for the isolated bases and GC assemblies, and
the PW91 functional \cite{perdew91} for the AT$_{\rm H}$ pair.  This latter
functional was widely tested in guanine-rich systems
\cite{calzolari1,calzolari2,Zhang_unpublished}, and yields a similar accuracy for
AT$_{\rm H}$ pairs and stacked pairs. 

It 
is well known that LDA may be grossly wrong in the description of H-bonds: 
that is why different gradient-corrected functionals were used to relax 
the geometries. However, we also know that, once the geometry is given, 
the the LDA and GGA furnish very similar results for the excitations, 
therefore we stick to the LDA functional for simplicity. We are confident this is
a reasonable approximation. Indeed, in the test mentioned above for guanine, 
different functionals do not yield significant differences as far as the 
optical excitations are concerned.

\section{Isolated gas-phase nucleobases}
\label{iso_bases}

In this section we present the calculated absorption spectra for the five
isolated nucleobases.  Both the purines and the pyrimidines exist in nature in
different tautomeric forms. We limited our calculations to the 9H keto form of
guanine, and to the 9H amino form of adenine. The latter tautomer is the one
present in DNA and RNA polymers, and is therefore the most relevant regarding
the adenine properties in nucleic acids.

\begin{figure}[!htb]
  \begin{center}
      \includegraphics[width=7cm]{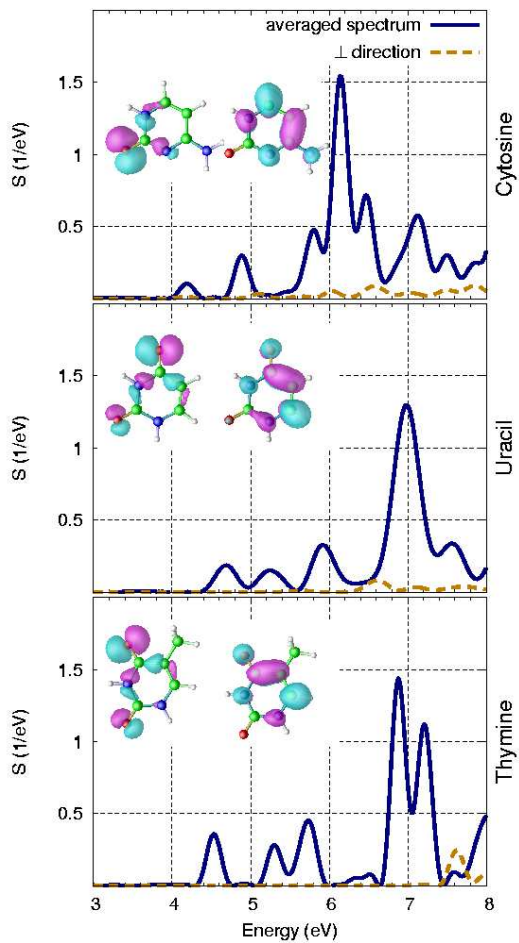}
    \caption[]{Photo-absorption cross section of isolated gas-phase pyrimidine
      nucleobases. The solid blue (dashed yellow) line is the signal averaged
      along the three real-space axes (projected onto the axis perpendicular to
      the base plane).  Insets: HOMO (left) and LUMO (right) Kohn-Sham wave
      functions.  The cyan (magenta) isosurfaces represent positive (negative)
      charge values.  Different atoms are indicated with different colors:
      carbon (green), nitrogen (blue), oxygen (red), hydrogen (white).}
    \label{basesa}
  \end{center}
\end{figure}

\begin{figure}[!htb]
  \begin{center}
      \includegraphics[width=7cm]{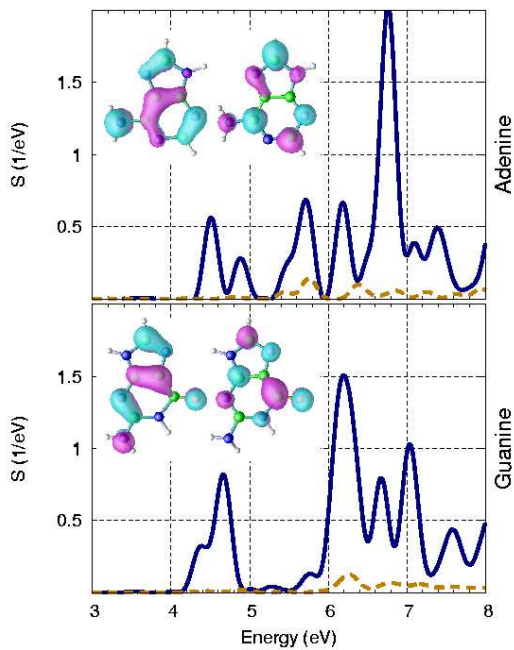}
    \caption[]{
      The same as Fig.~\ref{basesa} for adenine and guanine.
    }
    \label{basesb}
  \end{center}
\end{figure}

\subsection{Pyrimidines}

\subsubsection{Cytosine}

\begin{table}
  \begin{tabular}{ccccc}
    Exp.& TDDFT\textsuperscript{(a)} & TDDFT\textsuperscript{(b)} 
    & CASPT2\textsuperscript{(c)}& TDDFT\textsuperscript{(d)}\\
    \hline
    4.6     &4.18  &4.10& 4.39& 4.65 \\
    5.0-5.3 &4.88  &4.90& 5.36& 5.39 \\
    5.4-5.8 &5.80  &5.92&     & 6.11 \\
    6.1-6.3 &6.14  &6.39& 6.16& 6.32 \\
    &6.47  &6.48& 6.74& 6.46 \\
    6.7-7.1 &6.88  &6.88& 7.61&      \\
    &7.13  &7.16&     &      \\
    &7.50  &    &     &  
  \end{tabular}
  \caption{Vertical excitations energies (eV) calculated for cytosine,
    compared with averaged experimental values and selected computational
    results. The columns correspond to: (a)~this work;
    (b)~Ref.~\onlinecite{kaxiras}; (c)~Ref.~\onlinecite{fulscher95}; (d)~Ref.~\onlinecite{shukla04}. 
The experimental absorption and CD values, along with the original references, 
are collected in Ref.~\onlinecite{fulscher95}.}
  \label{tabcyt}
\end{table}

The computed total and perpendicular photo-absorption cross sections of cytosine
are shown in Fig.~\ref{basesa}. The energies of the spectral peaks are
reported in Table~\ref{tabcyt}, together with other selected sets of data that
represent the state-of-the-art performance of experimental and computational
techniques on cytosine.

For what concerns the transition energies, an overall glance at
Table~\ref{tabcyt} reveals a rather good agreement of our
results with those obtained in Ref.~\onlinecite{kaxiras}, and a poorer matching
with the outcome of B3LYP-TDDFT based on GAUSSIAN98 
\cite{shukla04}, and of CASPT2.  None of the computational data is in perfect
accord with experiment. Therefore, a selection of the best theoretical approach
on the basis of the comparison between theory and experiment is hindered. Our
agreement with experimental values is much more satisfactory in the high-energy
than in the low-energy regime.

If we turn our attention to the relative oscillator strengths between the
various peaks, then the agreement with the data of Ref.~\onlinecite{kaxiras} is
not so perfect: we encounter differences in the distribution of oscillation
strengths in the high-energy range of the spectrum. We will see later that these
high energy discrepancies also appear in the spectra of the other nucleobases
and can be traced to the limitations of the localized basis set used in
Ref.~\onlinecite{kaxiras}, and absent in the present work due to the use of a
real-space grid.

Combining the time-domain spectra with the frequency-domain analysis, we can
describe the spectral peaks in terms of transitions between electron states.
Fig.~\ref{fig:casida} shows an $n\pi^*$ transition at 4.3\,eV in the
perpendicular spectrum, not clearly discernible in Fig.~\ref{basesa} because of
the low intensity of the perpendicular signal. This means that the lowest energy
peak at 4.18\,eV in the averaged spectrum (Table~\ref{tabcyt}), absent in the
perpendicular polarization, must have a $\pi\pi^*$ character.  The fact that the
first optical excitation in cytosine is $\pi\pi^*$-like was predicted in all
recent computational studies. The spacing between the lowest $\pi\pi^*$
excitation and the $n\pi^*$ transition computed by us is 0.12\,eV: this value
is in good agreement with the results of B3LYP-TDDFT
\cite{shukla04} and CIS \cite{shukla99}, whereas CASPT2 calculations
\cite{fulscher95,merchan} predict a larger spacing between the two excitations.
A smaller spacing was found in Ref.~\onlinecite{ismail} with B3LYP-TDDFT.

\subsubsection{Uracil and thymine}

\begin{table}
  \begin{tabular}{cccc}
    Exp.& TDDFT\textsuperscript{(a)} & CASPT2\textsuperscript{(b)} & TDDFT\textsuperscript{(c)} \\
    \hline
    4.6-4.9      &4.69   & 4.54 $(n\pi^*)$ & 4.66 $(n\pi^*)$ \\
    4.9-5.2\textsuperscript{d}  &5.25 & 5.00 &  5.17 \\
    5.8-6.1      &5.92  &  5.82 & 5.89 \\
    6.3-6.6      &6.28  & 6.46 & 6.42 \\
    6.8-7.0      &6.98  & 7.00 & 6.81 \\   
    &7.57  & & 
  \end{tabular}
  \caption{Vertical excitations energies (eV) calculated for uracil, compared 
    with average experimental values and selected calculations. The columns
    correspond to: (a)~this work; (b)~Ref.~\onlinecite{lorentzon}; (c)~Ref.~\onlinecite{shukla04};
    (d)~this band was obtained in CD spectra and it was argued that it exhibits $n\pi*$ character.
The experimental absorption and CD values,
 along with the original references, are collected in Ref.~\onlinecite{lorentzon}; 
 the second experimental band that is taken from
Refs.~\onlinecite{miles,sprecher}.}
  \label{tabura}
\end{table}

\begin{table}
  \begin{tabular}{ccccc}
    Exp.& TDDFT\textsuperscript{(a)} &  TDDFT\textsuperscript{(b)} 
    & CASPT2\textsuperscript{(c)} & TDDFT\textsuperscript{(d)} \\
    \hline
    4.5-4.7      &4.54  &4.45  & 4.39$(n\pi^*)$ &4.69\\
    5.0-5.1      &5.30  &5.24 & 4.88 & 4.96 \\
    5.8-6.0      &5.74  &5.68 & 5.88&  5.95 \\
    6.3-6.6      &6.36  &6.38 &6.10 & 6.19 \\
    &6.51  &     &     & 6.50 \\
    7.0          &6.88  &6.86 & 7.13 & 6.86 \\
                 &      &7.07 & & \\
                 &7.21  &7.28 & & \\
                 &7.59  &7.52 &
  \end{tabular}
  \caption{Vertical excitations energies (eV) calculated for thymine, compared 
    with average experimental values and selected calculations. The columns
    correspond to: (a)~this work; (b)~Ref.~\onlinecite{kaxiras};
    (c)~Ref.~\onlinecite{lorentzon}; (d)~Ref~\onlinecite{shukla04}.
The experimental absorption and CD values, along with the original 
references, are collected in Ref.~\onlinecite{lorentzon}; 
the experimental second band that is taken from
Refs.~\onlinecite{miles,sprecher}.
} 
  \label{tabthy}
\end{table}

The computed total and perpendicular photoabsorption cross sections of uracil
and thymine are shown in Fig.~\ref{basesa}.  The energies of the spectral peaks
are reported in Tables~\ref{tabura} and~\ref{tabthy}.  The spectra of the two
bases U and T are very similar, both in the number of peaks and spectral
energies.  A fairly good agreement is met with averaged experimental
data, as well as with TDDFT \cite{kaxiras,shukla04} and CASPT2 \cite{lorentzon}
computational data.

All previous calculations, either in the TDDFT framework or with quantum
chemistry methods, predict the lowest transition to have an $n \pi^*$ character
in vacuo.  We agree with this assignment for uracil (the perpendicular signal is
not visible at 4.69\,eV in Fig.~\ref{basesa} due to the weak intensity).  In the
case of thymine, we do not find any appreciable signal perpendicular to the base
plane contributing to the peak at 4.54\,eV. However, this is consistent with
TDDFT \cite{shukla04} and CASPT2 \cite{lorentzon} calculations, as they indeed
report $n\pi^*$ transitions with extremely small oscillation strengths
(1e-4--1e-6), unresolved by us.

The character of the band at 4.9--5.2\,eV for uracil and at 5.0--5.1\,eV for
thymine is not yet clear: the absorption spectra of uracil compounds
\cite{clark65} and circular dichroism experiments \cite{miles} indicate that it is
due to a $\pi\pi^*$ transition. However, such $\pi\pi^*$ character has not been
confirmed by magnetic circular dichroism \cite{voelter} and polarized absorption
spectra experiments \cite{stewart63,stewart64,fucaloro,eaton}.  Lorentzon and coworkers, by
performing calculations at the CASPT2 level \cite{lorentzon}, suggest that this
band has a $n\pi^*$ origin: they arrived at this conclusion by correcting their
computed values with a 0.5\,eV blue-shift due to the solvent.  A similar
conclusion was reported for uracil by Shukla and coworkers \cite{shukla00}, who
performed CIS calculations taking the solvent into account with a polarizable
continuum model.  However, more recently, Shukla and Leszczynski using
B3LYP-TDDFT \cite{shukla04} reported only a $\pi\pi^*$ transition at 5.17\,eV
for uracil and at 4.96 for thymine in the energy range under consideration.  In
our work, we do not detect any absorption in the direction perpendicular to the
base-plane in this energy range, whereas peaks at 5.24\,eV for T and at 5.25\,eV
for U are found, with considerable oscillator strength induced by light
polarized in the plane of the molecule. Therefore, our results indicate that the
second band most likely has a $\pi\pi^*$ character.

\subsection{Purines}

\subsubsection{Adenine}

\begin{table}
  \begin{tabular}{ccccc}
    Exp.& TDDFT \textsuperscript{(a)} &  TDDFT\textsuperscript{(b)} &
    CASPT2\textsuperscript{(c)} & TDDFT\textsuperscript{(d)} \\
    \hline
    4.59    & 4.51 & 4.51 &5.13&4.94   \\
    4.8-4.9 & 4.88 & 4.95 &5.20&5.21   \\
    5.38    & 5.49 & 5.58 &    &       \\
    5.7-6.1 & 5.72 & 5.79 &6.24& 5.93  \\
    6.26    & 6.19 & 6.28 &6.72& 6.12  \\
            & 6.49 & 6.63 &    & 6.16 \\
    6.81    & 6.76 & 6.92 &6.99&\\
            & 7.09 & 7.47 &    &\\
    7.73    & 7.39 & 7.81 &7.57&
  \end{tabular}
  \caption{Vertical excitations energies (eV) calculated for adenine, compared 
    with averaged experimental values and selected calculations. The columns
    correspond to: (a)~this work; (b)~Ref.~\onlinecite{kaxiras};
    (c)~Ref.~\onlinecite{fulscher97}; (d)~Ref.~\onlinecite{shukla04}.
The experimental absorption and CD values, along with the original 
references, are collected in Ref.~\onlinecite{fulscher97}.}
  \label{tabade}
\end{table}

The computed total and perpendicular photo-absorption cross sections of
9H-adenine are shown in Fig.~\ref{basesb}.  The 
spectral peaks are summarized in Table~\ref{tabade} and compared with other
available theoretical and experimental data.  An excellent agreement with other
LDA-TDDFT computed values \cite{kaxiras} is observed, regarding both the peak
energies and the relative oscillation strengths.  Slight discrepancies occur
only in the high energy range: these are most likely due to differences in
the technical details adopted in the calculations (e.g., the basis sets).

Looking at the spectrum in the direction perpendicular to the base plane, we can
distinguish four prevalently $n\pi^*$ transitions in the energy range
4.10--4.79\,eV, that are hidden in Fig.~\ref{basesb} because of their tiny
oscillator strengths.  We find that the lowest excitation has mainly $n\pi^*$
character and is very close in energy to the first $\pi\pi^*$ transition
(forming the peak at 4.51\,eV).  This result is in agreement with resonant
two-photon ionization and laser induced fluorescence spectroscopy of jet cooled
adenine \cite{kim}, provided that the adiabatic transition energies follow the
same trends as the calculated vertical ones.  Our findings also agree
qualitatively with frequency-domain B3LYP-TDDFT \cite{shukla04} and CIPSI
\cite{mennucci,sobolewski02} calculations, whereas CASPT2 \cite{fulscher97} and CIS
\cite{holmen,broo,mishra} yield the reverse order. Given the small energy
difference between the two transitions, we can conclude that the overall
qualitative agreement is satisfactory.

Proceeding to higher energies, we find the second $\pi\pi^{*}$ peak 0.37\,eV
after the first one.  Experiments indicate \cite{holmen,callis,clark95} that the
low energy portion of the adenine photo-absorption spectrum consists of two
closely spaced $\pi\pi^*$ transitions.  Holm\'en and coworkers showed by linear
dichroism that those two peaks are separated by 0.26\,eV \cite{holmen}. This
value is in reasonable agreement with our results and with other TDDFT
calculations \cite{mennucci,kaxiras}, while 
different techniques predict a closer energy spacing.  On the other hand,
measured spectra indicate that the lowest energy transition has a smaller
oscillator strength than the subsequent one: such an evidence is consistent with
the CASPT2 and CIPSI results, whereas TDDFT simulations yield an inverse
ordering of oscillator strengths in the first two peaks \cite{kaxiras} (see
Fig.~\ref{basesb}).

Concerning other $n\pi^*$ transitions, there is experimental evidence by
circular \cite{sprecher,brunner} and linear \cite{holmen} dichroism of a signal
around 5.4\,eV. Our calculation is also able to reveal a signal in this range.
However, the agreement with the experiment for the $n\pi^*$ transitions should
be considered with caution, because of their high sensibility to the effect of
the solvent (neglected in our calculations) and exchange-correlation functional.

\subsubsection{Guanine}

\begin{table}
  \begin{tabular}{ccccc}
    Exp.& TDDFT \textsuperscript{a} &  TDDFT\textsuperscript{b} & CASPT2\textsuperscript{c} &TDDFT\textsuperscript{d}\\
    \hline
    4.4-4.5 & 4.40 & 4.46 &4.76&4.85 \\
    4.9-5.0 & 4.66 & 4.71 &5.09&5.11 \\
            & 5.01 & 5.04 &    &     \\
            & 5.28 &      &    &     \\
    5.7-5.8 & 5.76 & 5.64 &5.96&5.59 \\
    6.1-6.3 & 6.22 & 6.23 &6.55&5.83 \\
    6.6-6.7 & 6.67 & 6.53 &6.65& \\
            &      &      &6.66& \\
            &      & 6.82 &6.77& \\
            & 7.04 & 6.93 &    & \\
            & 7.58 & 7.26 &    & 
  \end{tabular}
  \caption{Vertical excitations energies (eV) calculated for guanine, compared 
    with averaged experimental values and selected calculations. The columns
    correspond to: (a)~this work; (b)~Ref.~\onlinecite{kaxiras};
    (c)~Ref.~\onlinecite{fulscher97}; (d)~Ref.~\onlinecite{shukla04}.
The experimental absorption and CD values, along with the original 
references, are collected in Ref.~\onlinecite{fulscher97}.}
  \label{tabgua}
\end{table}

To conclude the presentation of the results for the isolated DNA bases, we show
in Fig.~\ref{basesb} the computed total and perpendicular photo-absorption cross
sections of 9H-guanine. The spectral peaks are summarized in
Table~\ref{tabgua} and compared with other available data.  The excitation
energies are in good agreement with the averaged experimental data and with the
results of Ref.~\onlinecite{kaxiras}. Small differences are encountered in the
high energy region of the spectrum in the relative oscillator strengths of the
peaks at 6.22\,eV and 6.67\,eV, while the low energy region is in perfect
agreement. Again, we attribute the discrepancies at higher energies to the different basis
sets used in Ref.~\onlinecite{kaxiras} and in our work. 
In the region from 5 to 6\,eV we also
find very weak peaks that are not observed experimentally.
  
Regarding the out-of-plane spectrum, we find one peak at 4.47\,eV that has a
purely $n\pi^*$ character. Thus, the first $n\pi^{*}$ transition has a higher
energy than the first $\pi\pi^*$ transition found at 4.40\,eV (see
Table~\ref{tabgua}).  This attribution is consistent with previous CASPT2
\cite{fulscher97}, CIS \cite{shukla00}, and B3LYP-TDDFT \cite{crespo_unp}
calculations.

\subsection{General comments on the results for the isolated gas-phase nucleobases}

The results of our time-dependent calculations~\cite{octopus}
of the excitation spectra of DNA bases and uracil in the gas
phase show a satisfactory agreement with the experimental data and with other
computational approaches, especially for the most intense peak above 6\,eV.  In
particular, good agreement is found with recent LDA-TDDFT calculations using a
localized-basis set\cite{kaxiras}.  Minor differences are found regarding the
distribution of the oscillator strengths in the high energy range of the
spectra.  Such discrepancies can be ascribed to the different basis sets: 
a uniform grid in real space in this work, and a set of
localized atomic orbitals in Ref.~\onlinecite{kaxiras}.

Regarding the more problematic $n\pi^*$ transitions, we find that the weak
perpendicular signal is sometimes hidden below the in-plane signal.  In the
cases where the $n\pi^*$ transitions can be detected, we find the correct
relative ordering between $n\pi^*$ and $\pi\pi^*$ peaks, as compared with other
TDDFT calculations performed in the frequency domain.

Another interesting trend is the relative ordering among different nucleobases
for what concerns the first $\pi\pi^*$ transition.  For this feature, supersonic
jet experiments report $C < G < A$ \cite{kim,mons,nir}.  We indeed find that
ordering in our first-principles calculations, i.e.  the first excitation at
4.18\,eV for cytosine, 4.40\,eV for guanine and at 4.52\,eV for adenine.

Last, we note that our calculations give the expected character of the
Kohn-Sham orbitals of the different bases (see insets in Figs.~\ref{basesa}
and~\ref{basesb}): the LUMO has $\pi$ character for all the nucleobases, whereas
the HOMO is $\pi$-like for the purines and $\sigma$-like for the pyrimidines,
with a high charge density around the oxygen atoms in all cases.

\section{Watson-Crick pairs GC$_{\rm H}$ and AT$_{\rm H}$}
\label{wats_crick}

\begin{figure*}[!tb]
  \begin{center}
    \begin{tabular}{cc}
    \includegraphics[width=7cm]{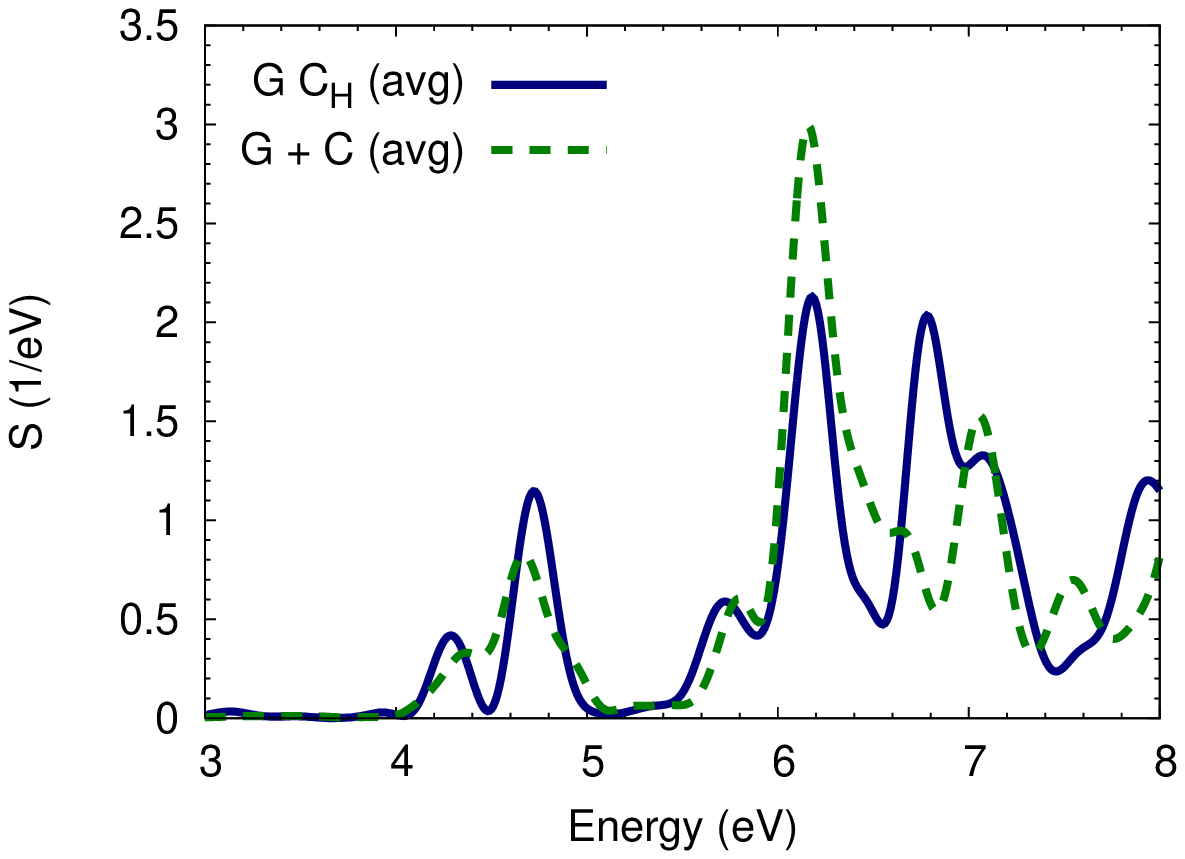}  & 
      \includegraphics[width=7cm]{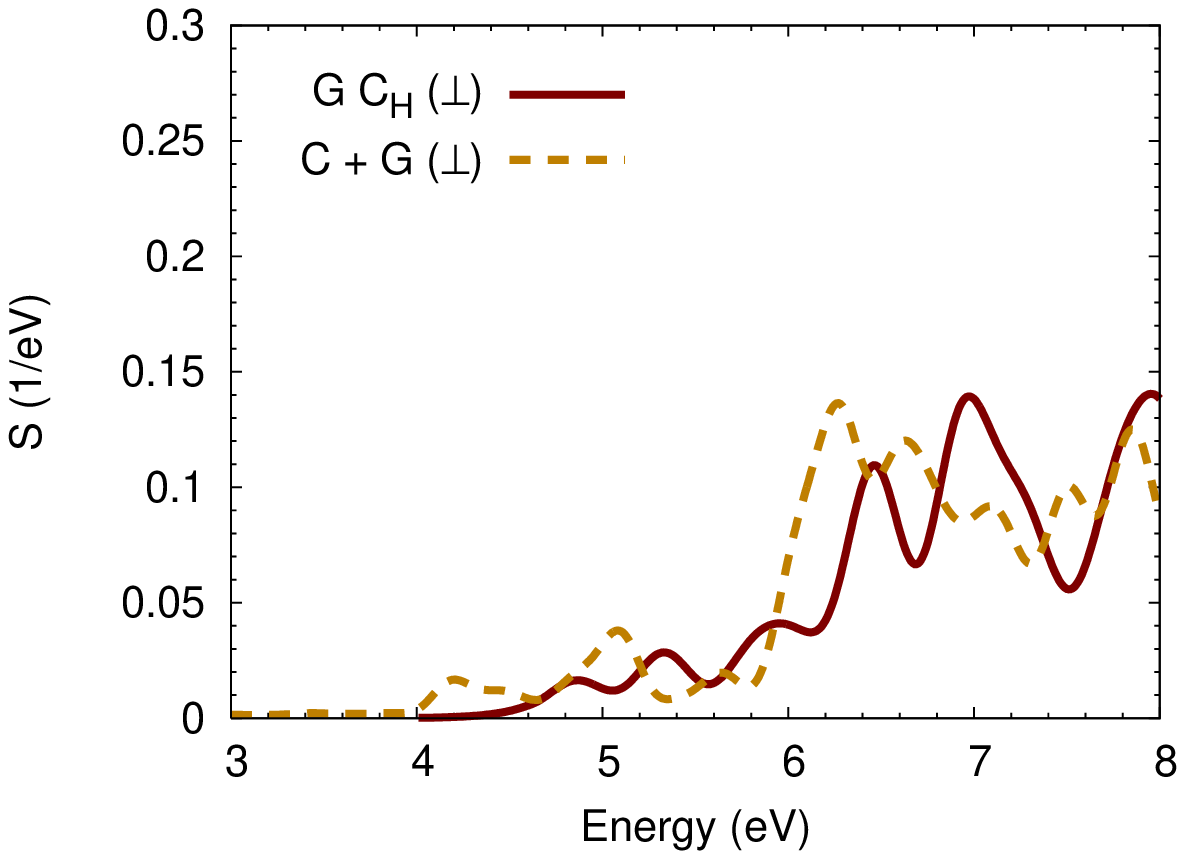}  \\
     & \\
      \includegraphics[width=7cm]{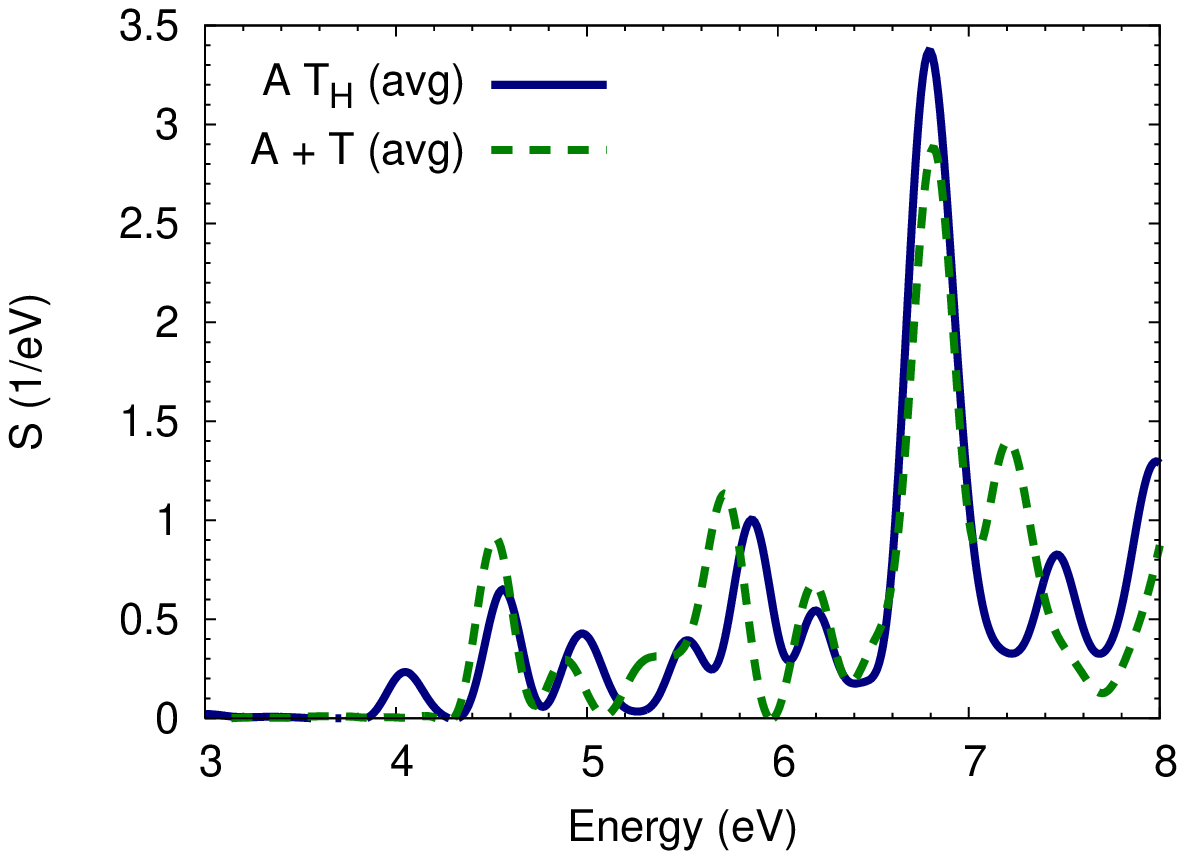}  & 
      \includegraphics[width=7cm]{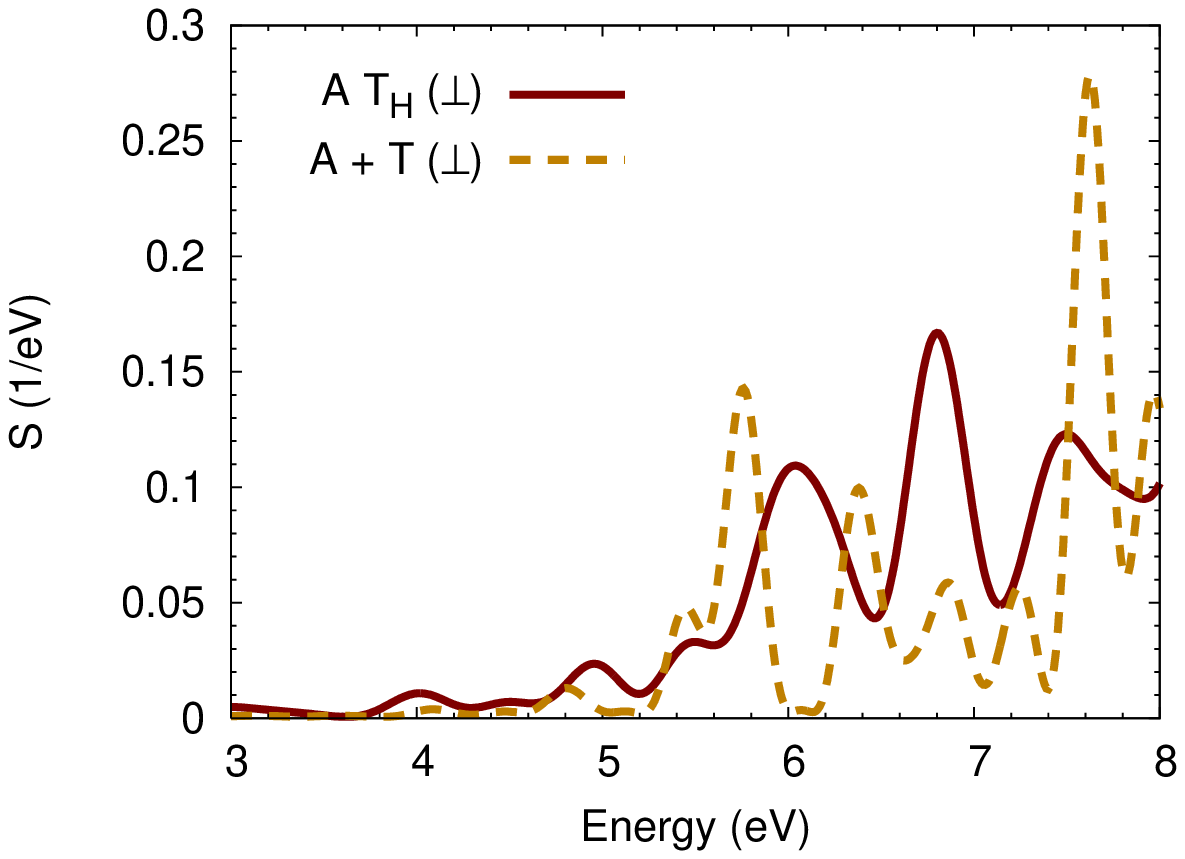}  \\
    \end{tabular}
    \caption[]{Photo-absorption cross section of the GC$_{\rm H}$ 
      and AT$_{\rm H}$ base pairs averaged (solid blue, left) and in the direction
      perpendicular to the base plane (solid red, right). The linear combinations of
      the spectra of the isolated purine and corresponding pyrimidine (G+C and
      A+T; dashed lines) are also shown for comparison.}
    \label{hbond}
  \end{center}
\end{figure*}

\begin{table}
  \begin{tabular}{ccc|cc}
    \multicolumn{3}{|c|} {GC} & \multicolumn{2}{|c|}{AT} \\   
      
    TDDFT\textsuperscript{(a)} &  TDDFT\textsuperscript{(b)} &
    CASPT2\textsuperscript{(c)} & TDDFT \textsuperscript{(a)} & TDDFT\textsuperscript{(b)} \\
    \hline
    2.44    & 2.37 &      &     &   \\
    3.14   & 3.53 &      &    &3.22   \\
    3.94   &      &       & 4.05   & 4.05      \\
    4.29 & 4.21 & 4.35 &     &4.26  \\
    4.72 & 4.75 & 4.67 &4.56 &4.40  \\
         &      & 4.75 \textsuperscript{(d)} &4.98  &4.80 \\
    5.38 & 5.52 &      &     &5.20 \\
    5.72 & 5.69 &      &     &5.45 \\
         & 5.84 &      &5.53 &5.60 \\
         & 6.07 &      &5.87 &5.76 \\
    6.17 & 6.33 &      &6.20 &6.24 \\
    6.46 & 6.47 &      &     &6.48 \\
    6.78 & 6.68 &      &6.78 &6.73 \\
         & 6.83 &      &7.80 &6.97 \\
    7.07 & 7.18 &      &     &     \\
    7.53 & 7.28 &      &     &     \\
  \end{tabular}
  \caption{Vertical excitation energies (eV) for Watson-Crick GC$_{\rm H}$ and 
    AT$_{\rm H}$ pairs, compared to other calculations. The columns correspond to:
    (a)~this work; (b)~Ref.~\onlinecite{kaxiras};
    (c)~Ref.~\onlinecite{sobolewski04} (cf. also the results of Shukla and Leszczynski\cite{shukla02});
    (d) Charge transfer transition.}
  \label{tabwc}
\end{table}

\begin{figure}[!tb]
  \begin{center}
    \begin{tabular}{ccc}
      & {\bf HOMO}  & {\bf LUMO}  \\[0.2cm]
      \hspace{0.5cm}{\bf GC$_{\rm H}$}
      & \includegraphics[width=3.5cm]{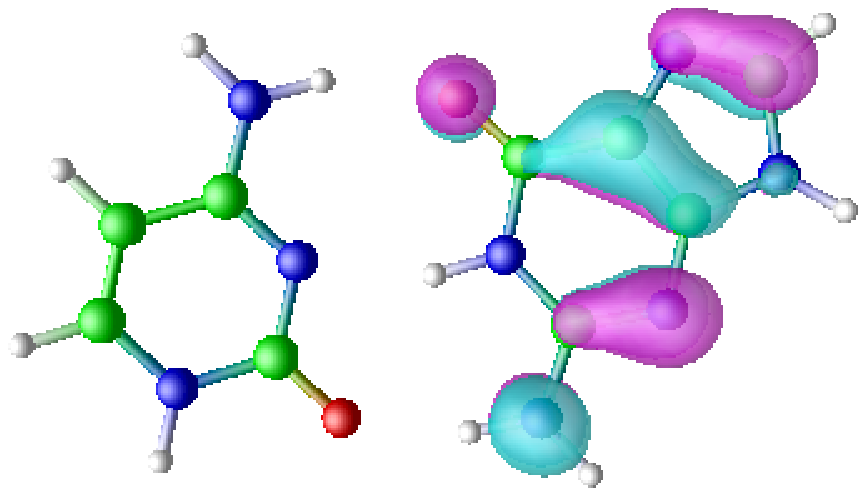} \hspace{0.5cm} 
      & \includegraphics[width=3.5cm]{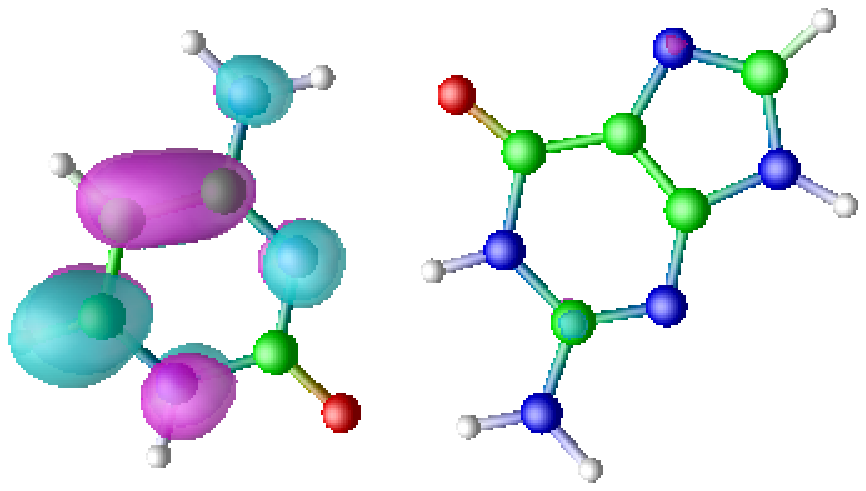}  \\[0.5cm]
      \hspace{0.5cm}{\bf AT$_{\rm H}$} 
      & \includegraphics[width=3.5cm]{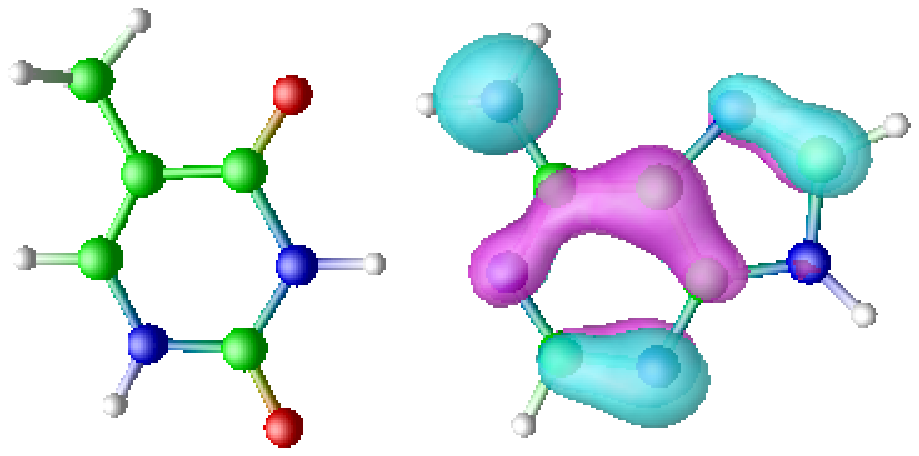} \hspace{0.5cm}
      & \includegraphics[width=3.5cm]{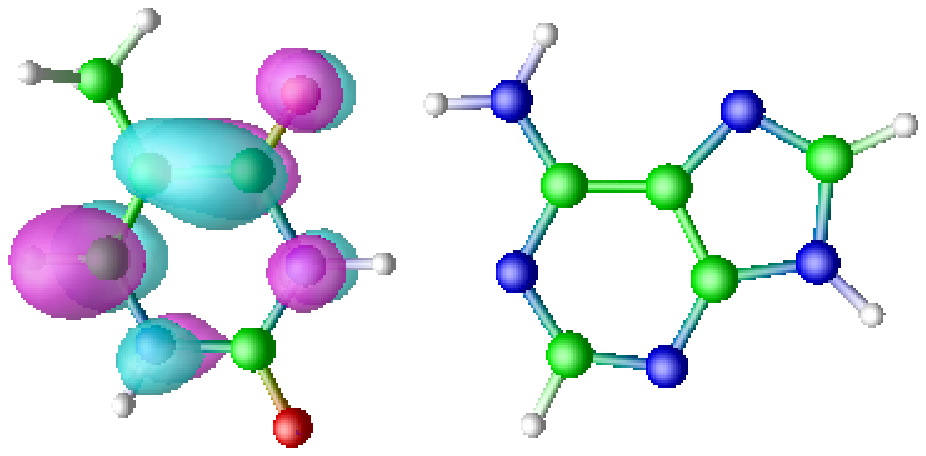}
    \end{tabular}
    \caption[]{HOMO and LUMO Kohn-Sham wave functions of the 
      GC$_{\rm H}$ and AT$_{\rm H}$ pairs. }
    \label{homolumohbond}
  \end{center}
\end{figure}

We now turn the discussion to the standard assembly of DNA bases, namely the
Watson-Crick pairing.  Figure~\ref{hbond} shows the calculated TDDFT spectra of
H-bonded GC$_{\rm H}$ and AT$_{\rm H}$ pairs (left) and the spectra resolved in
the direction perpendicular to the basal plane (right).  The sum of the
photo-absorption cross sections from the isolated component nucleobases is also
shown by the dashed lines (indicated by G+C and A+T), to reveal the effects induced by H-bonding in
the Watson-Crick pairs.  The peaks observed in the spectra are also listed in
Table~\ref{tabwc} for a more detailed analysis.  Figure~\ref{homolumohbond}
illustrates the isosurface plots of the HOMO and LUMO, which are the
single-particle orbitals more important for the optical transitions in the
energy range discussed here. In both pairs, the HOMO is localized at the purine
and the LUMO at the pyrimidine. 
The HOMO-LUMO gaps are 1.93 eV and 3.07 eV 
for the GC$_{\rm H}$ and AT$_{\rm H}$ pairs, respectively.

We focus our analysis on the most evident features induced by H-bonding base
pairing, that are: (i)~a small shift found for the lowest frequency peaks;
(ii)~the hypochromicity (intensity decrease) at high frequencies; (iii)~an
overall blue-shift of the spectrum for light polarized perpendicular to the
pair.

The relaxed structure of the GC$_{\rm H}$ pair is practically planar.  The shape
of the total photo-absorption spectrum of the hydrogen-bonded GC$_{\rm H}$ pair
remains rather similar to the linear superposition of the isolated bases (cf.
the top left panel of Fig.~\ref{hbond}).  The first $\pi\pi^*$ peak at 4.29\,eV
is composed of the first excitation of cytosine and the first excitation of
guanine.  We note a slight redshift of this peak with respect to the first peak
in the red curve: this effect is the combination of a blue-shift by about
0.1\,eV of the first cytosine peak and a comparable redshift of the first
guanine peak.  Such finding is in qualitative agreement with a CASPT2
calculation by Sobolewski and Domcke \cite{sobolewski04} and by a calculation
performed by Wesolowski \cite{wesolowski} by embedding methods.

For the second $\pi\pi^*$ transition at 4.72\,eV the situation is reversed: The
peak is blue-shifted with respect to the superposition of isolated G and C. This
shift results from a blue-shift of the second guanine peak and a red-shift of
the second cytosine peak.
In the high-energy spectral range, we remark that the effect of the H-bonding pairing does
not change the position of the brightest peak at 6.17\,eV, which has both G and C
components.  However, the intensity of this peak is depressed (hypochromism) by
about 30\% with respect to the free monomers.  Above this energy the spectrum
results much more changed.

Looking at the direction perpendicular to the plane of the GC$_{\rm H}$ pair
(top right panel), we observe an overall blue-shift of the spectrum. In
particular, the first peak is shifted upward by 0.48\,eV. This behavior reflects
the well known uplifting of $n\pi^*$ excitation energies in hydrogen-bonding
environments, already pointed out by other authors \cite{callis83,brealey}, and
appears naturally from our first-principles simulation.

For the AT$_{\rm H}$ pair, Table~\ref{tabwc} highlights an overall good
agreement with a recent LDA-TDDFT calculation \cite{kaxiras}. Small differences
can be reasonably imputed to the different structure obtained upon relaxation.
We find an almost planar geometry, in agreement with Shukla and Leszczynski
\cite{shukla02}, while Tsolakidis and Kaxiras \cite{kaxiras} find a propeller
structure.

As in the case of the GC$_{\rm H}$ pair, the overall spectrum of AT$_{\rm H}$ is
quite similar to the linear combination of the spectra of the isolated A and T
bases.  A small blue-shift, by $\sim$0.1\,eV, is detected in the low-energy
range (roughly below 6\,eV) with respect the superposition of isolated bases.
In the high-energy part of the spectrum, we observe an enhancement of the
strength with respect to isolated adenine and thymine for the peak located at
6.78\,eV, and a reduction for the peak at 7.18\,eV present in the isolated
thymine spectrum. Hence, for the most intense peak at 6.78\,eV, the effect of
H-bonding is contrary to the hypochromicity described above for the GC$_{\rm H}$
pair.

In the out-of-plane spectrum (bottom right panel of Fig.~\ref{hbond}), as for
the GC$_{\rm H}$ pair, we observe that the effect of hydrogen bonding is an
overall blue-shift. In particular, the first peak (that has an $n\pi^*$
character) is shifted by 0.4\,eV; the second and fourth peaks are upward 
shifted by 0.2\,eV and 0.3\,eV respectively; the energy of the third peak 
is practically unchanged.

In addition to the above features that originate from the individual bases that enter 
each pair, new features appear in the spectra because the purine and the pyrimidine 
coexist in the Watson-Crick arrangement. 
This coexistence changes the nature of the frontier orbitals: 
The HOMO is purine-localized, the LUMO is pyrimidine-localized, and 
the value of the HOMO-LUMO gap is smaller than in the isolated bases. Consequently, 
peaks at lower energies emerge in the spectra: they are reported in Table VI, 
but the corresponding energy range is not shown in the figures, because 
the tiny 
oscillator strengths make them undiscernible from the more intense peaks. We 
also note that the extremely small intensity makes the quantitative assignment 
of these features less reliable than the others already discussed. 
In the case 
of GC$_{\rm H}$, we find excitation energies at 2.44 eV, 3.14 eV and 3.94 eV, 
with oscillator strengths of 0.001, 0.008 and 0.003. 
These oscillator
strengths are two orders of magnitude smaller than the peak at 4.29\,eV 
shown in the upper panel of Fig.~\ref{hbond}. 
For the AT$_{\rm H}$ pair, the lowest energy 
peak at 4.05\,eV has an oscillator strength of 0.048. 
This transition has a small component also in the
absorption for light polarized perpendicular to the base plane
(see bottom right panel in Fig.~\ref{hbond}). Its origin 
can be likely ascribed to charge transfer states, that have received particular 
attention from both experimental and theoretical viewpoints \cite{jortner}. 
However, we warn the reader that the identification of charge transfer transitions 
is to be taken with care at the current level of TDDFT when a local or 
gradient-corrected exchange-correlation functional is used~\cite{dreuw,dreuw2,gritsenko}.

\section{Stacked GC$_{s}$ and d(GC) structures}
\label{stacking}

\begin{figure*}[!tb]
  \begin{center}
    \begin{tabular}{cc}
      \includegraphics[width=7cm]{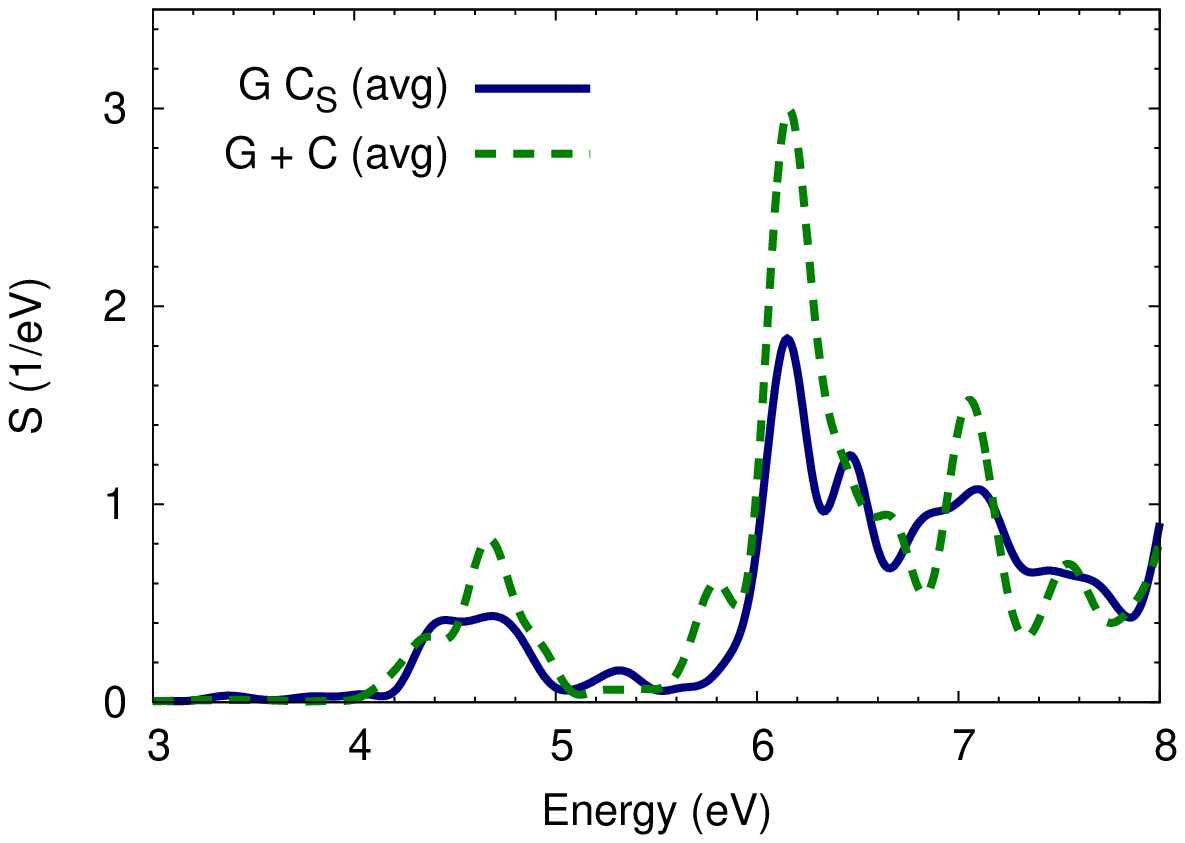}  & 
      \includegraphics[width=7cm]{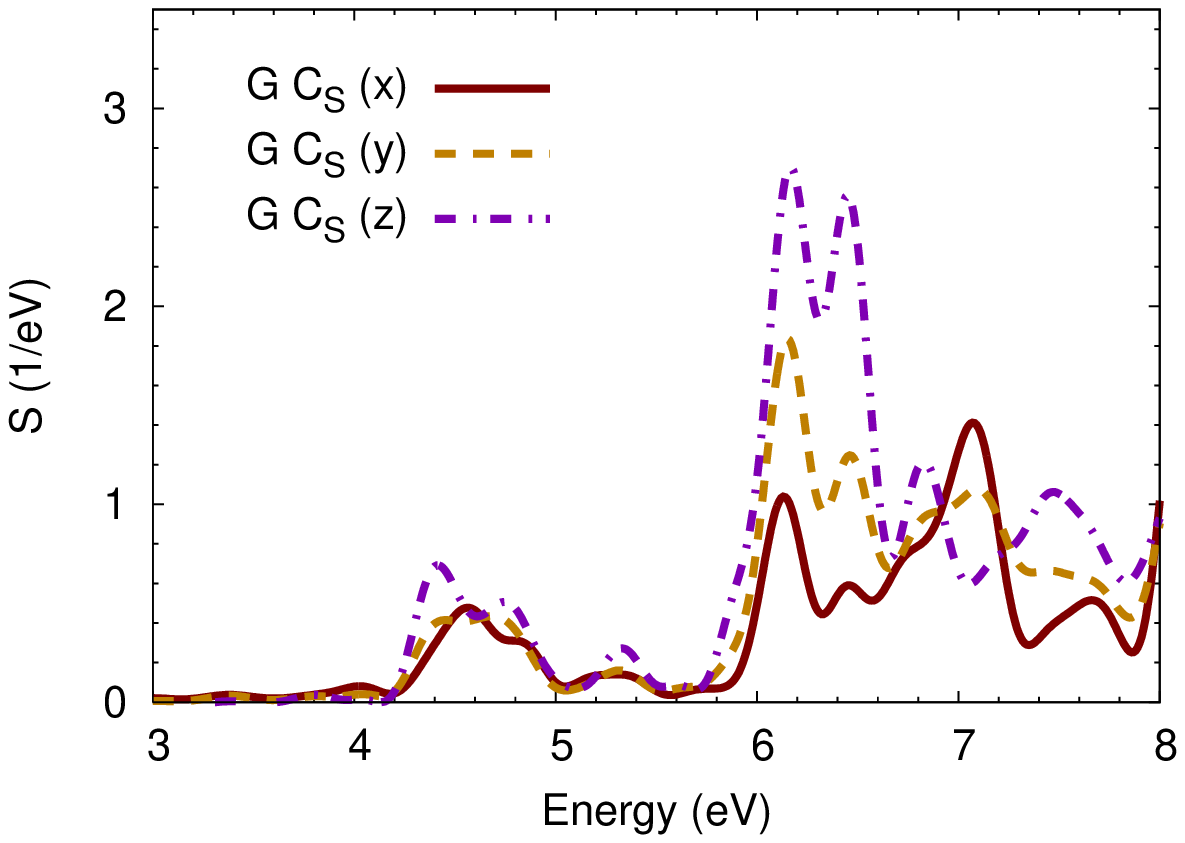}  \\
      \includegraphics[width=7cm]{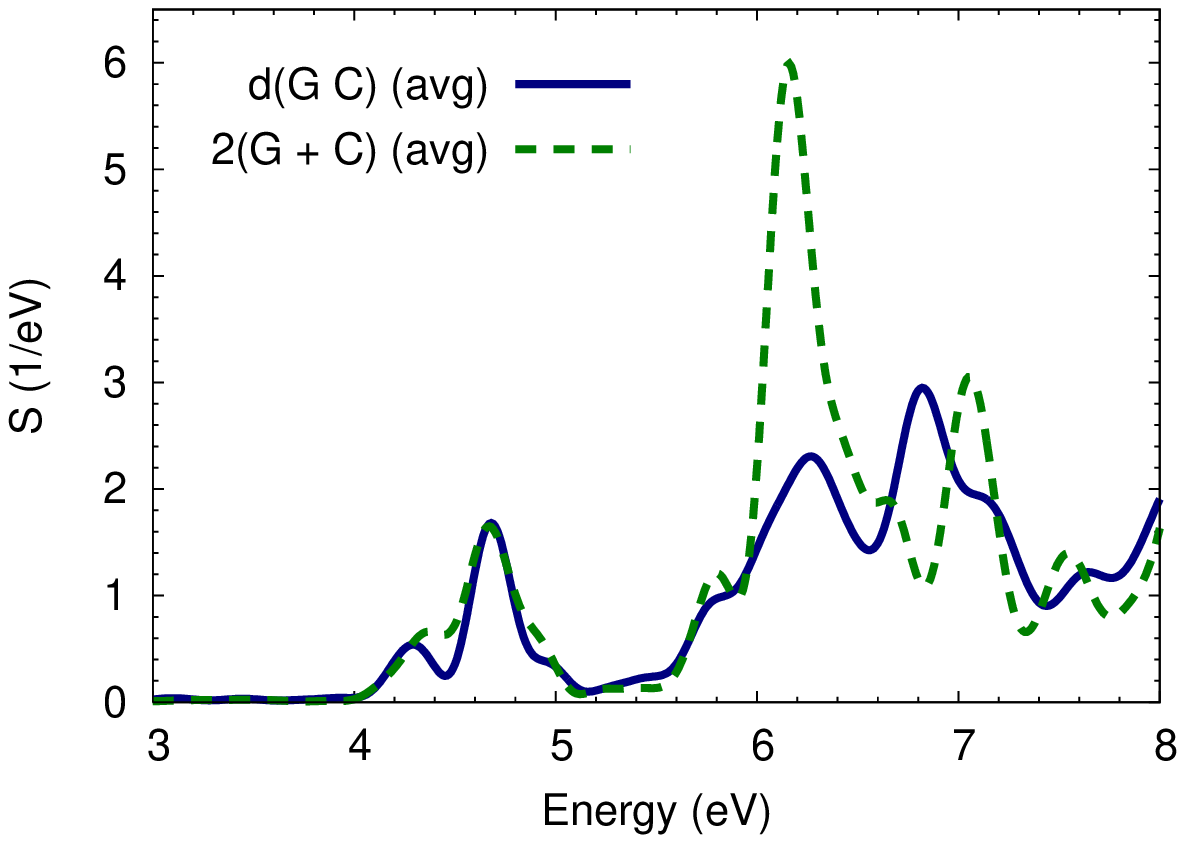}  & 
      \includegraphics[width=7cm]{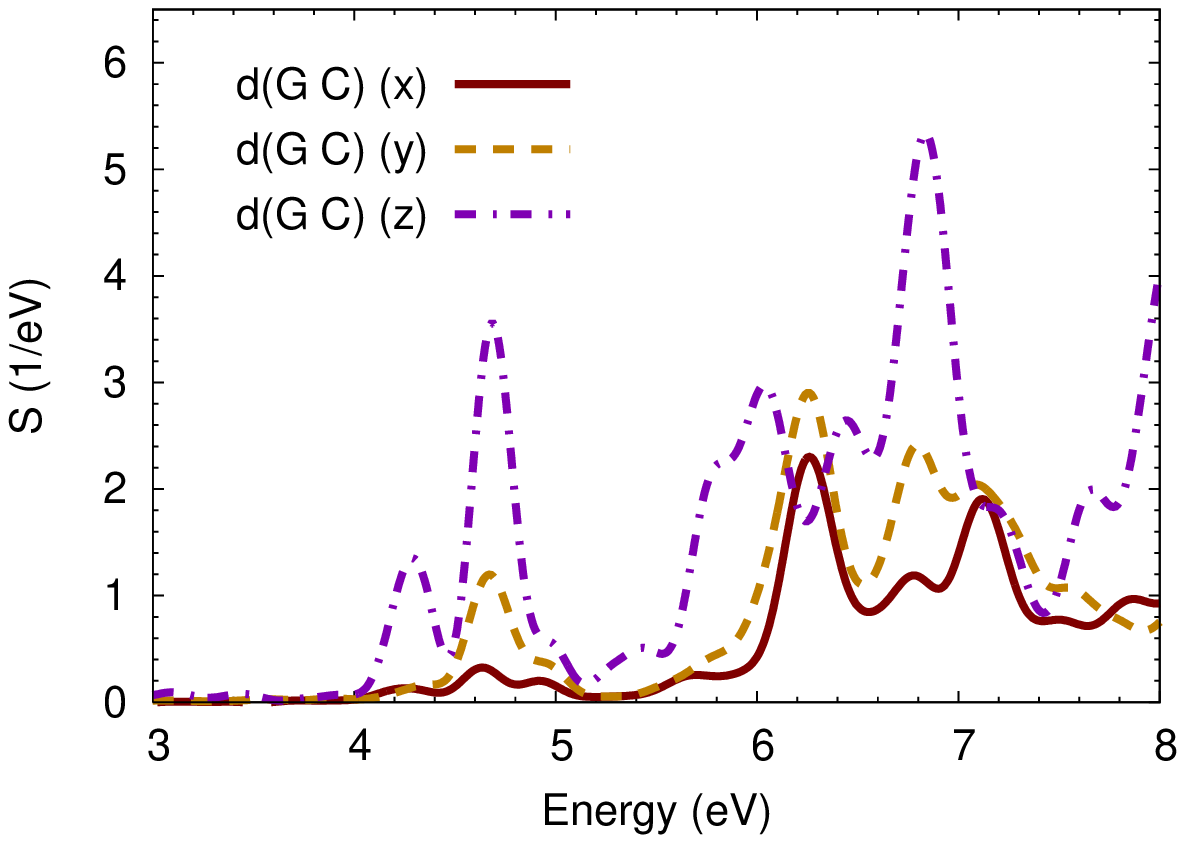}  \\
    \end{tabular}
    \caption[]{Photo-absorption cross section of the GC$_{\rm s}$ 
      stacked dimer (up) and of the d(GC) quartet (down). In each panel, the
      linear combination of the spectra of the two (up) and four (down)
      constituent bases is also shown. In the right panels the spectra resolved
      in the three spatial directions are shown.}
    \label{stack}
  \end{center}
\end{figure*}

\begin{figure}[!tb]
  \begin{center}
    \begin{tabular}{ccc}
      &  {\bf HOMO} & {\bf LUMO}  \\
      {\bf $\pi$ stacked base}
      & \includegraphics[width=2.5cm]{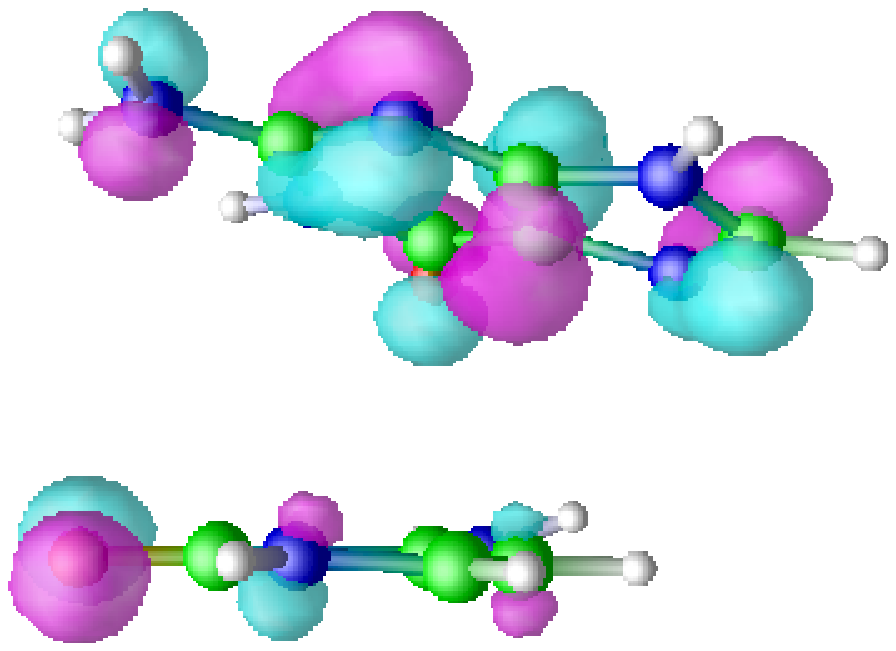}
      & \includegraphics[width=2.5cm]{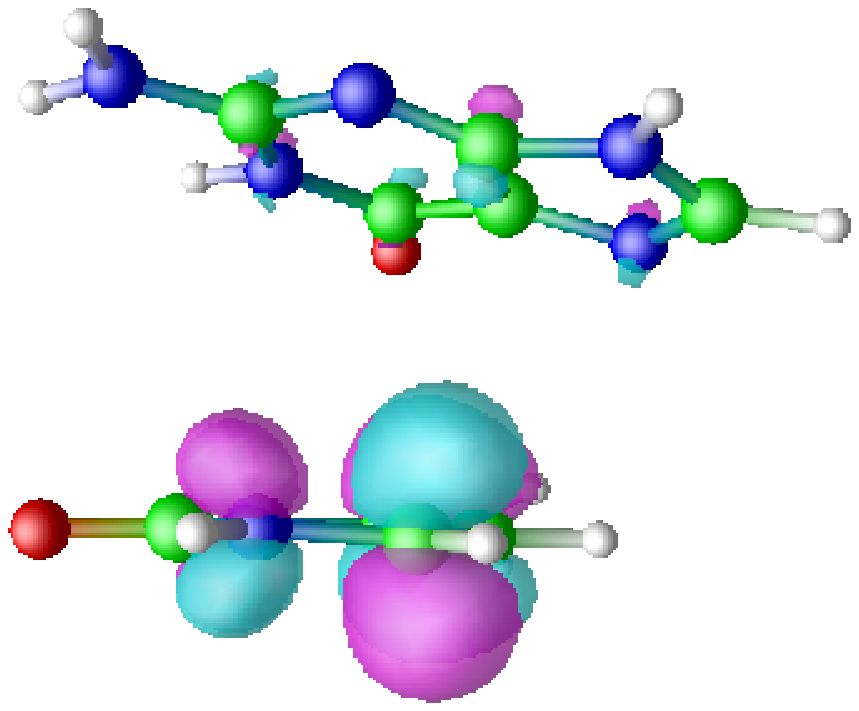}  \\[0.5cm]
      {\bf $\pi$ stacked pairs}
      & \includegraphics[width=3.8cm]{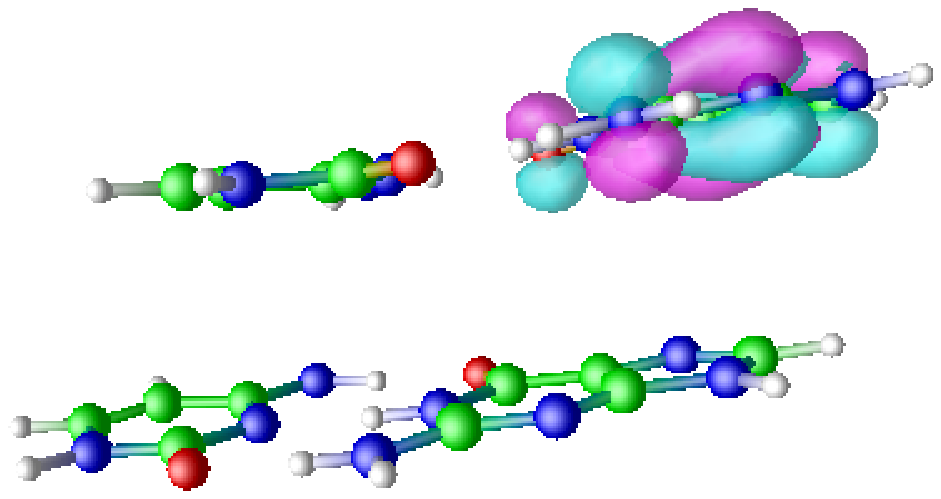}
      & \includegraphics[width=3.8cm]{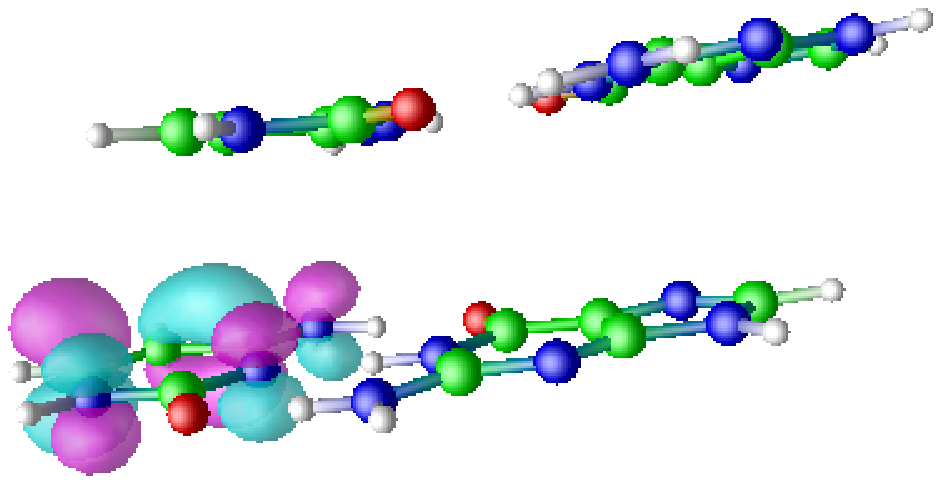} \\
    \end{tabular}
    \caption[]{Isosurface plots of the HOMO and LUMO Kohn-Sham wave 
      functions of the GC$_{\rm s}$ $\pi$-stacked dimer and of the d(GC)
      $\pi$-stacked quartet.}
    \label{homolumostack}
  \end{center}
\end{figure}

\begin{figure}[!tb]
  \begin{center}
    \begin{tabular}{c c}
      {\bf PDB 400d+CHARMM+B3LYP}  & {\bf Ref.~\onlinecite{difelice02} + PW91}  \\[0.5cm]
      \includegraphics[width=3.7cm]{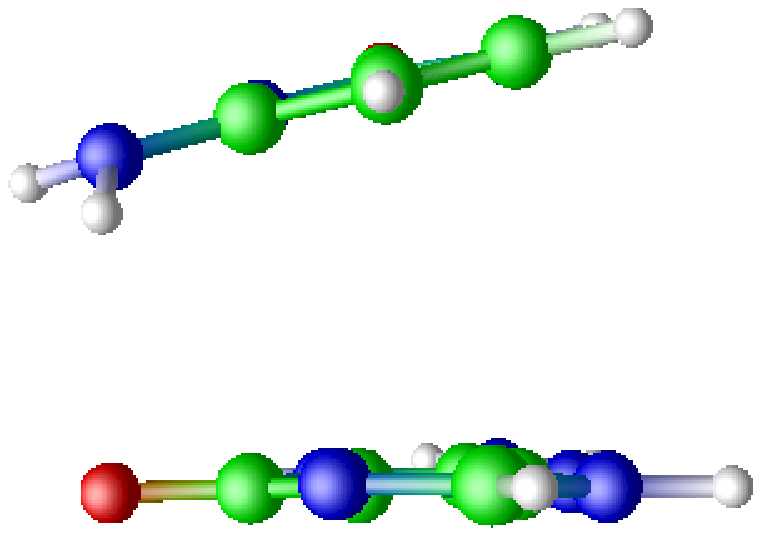}  &           
      \includegraphics[width=3.5cm]{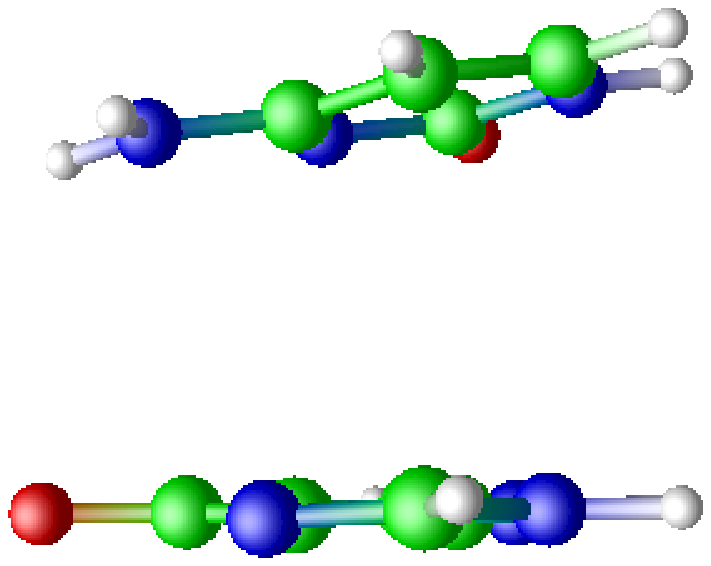}  \\
    \end{tabular}
    \caption[]{Two different conformations obtained for the $\pi$-staked GC$_{\rm s}$ 
      pair obtained by different structural relaxation schemes.}
    \label{geo}
  \end{center}
\end{figure}

With the results reported in the previous section for the GC$_{\rm H}$ pair, we
can understand what happens in the optical absorption when a guanine and a
cytosine dimerize through hydrogen bonding. Here, the GC$_{\rm s}$ system is
selected to check which effects arise when the same two bases dimerize in a
different form, namely by stacking.

Figure~\ref{stack} reports the averaged and direction-resolved spectra of the
stacked GC$_{\rm s}$ dimer. The comparison between Figs.~\ref{stack}
and~\ref{hbond} aims at disclosing the relative role of hydrogen-bonding and
$\pi$-stacking in the optical response of nucleobase complexes.  We also plot in
Fig.~\ref{stack} the linear combination of the spectra of the isolated
constituent bases.  Figure~\ref{homolumostack} (top) illustrates the Kohn-Sham
highest occupied and lowest unoccupied orbitals of the system.  It is important
to note that both the HOMO and the LUMO have charge density distributed around
both the purine and the pyrimidine, at odds with the H-bonded GC$_{\rm H}$ pair
(compare Fig.~\ref{homolumohbond}, top) where the HOMO is centered at the purine
and the LUMO at the pyrimidine.

As in the case of hydrogen-bonding, the shape of the spectrum it is not strongly
altered, but important differences are encountered in the oscillator strengths.
Looking at the low-energy peaks, we see that the excitations due to the
cytosine moiety are slightly blue-shifted (as for the GC$_{\rm H}$ pair),
while the guanine excitations appear unaffected. The resulting spectrum
below 5\,eV is made of two equally intense closely spaced peaks at 4.45\,eV and
4.68\,eV, whereas in the H-bonded GC$_{\rm H}$ structure we observe two well
separated peaks spaced by 0.4\,eV and with different intensities.  Hypochromicity
is also observed in this low-energy range, whereas for the GC$_{\rm H}$ pair
hypochromicity occurs only in the high-energy range.

In the range between 5 and 6\,eV, we find an enhancement of the weak
peak at 5.31\,eV and a depression of the peak at 5.78\,eV, which is present both
in the isolated bases (guanine and cytosine) and in the H-bonded pair.  In the
high-energy spectral range, the dominant effect is the 39\% hypochromicity for
the main peak located at 6.15\,eV: hypochromicity from $\pi$-stacking is thus
enhanced with respect to hypochromicity from hydrogen-bonding (30\%).  Above the
strongest peak, we find other less intense peaks at 6.47\,eV, 6.87\,eV, and
7.10\,eV. Instead, in the H-bonded GC$_{\rm H}$ pair, the high-energy range
contains a peak at 6.78\,eV with the same intensity as the peak at 6.17\,eV, and
then only an additional weaker one.

The calculation of the GC$_{\rm s}$ has been repeated for a geometry
obtained by a slightly different relaxation scheme, within the
 DFT-PW91 framework \cite{calzolari1,calzolari2}.
This additional GC$_{\rm s}$ geometry was considered
just for a gross check of structural deformations induced by computational
details. The starting configuration was fixed by taking guanine and cytosine
from a previous DFT study of isolated bases \cite{difelice02} and putting them
parallel to each other at an initial distance of 3.4 \AA.  The major difference
between the two GC$_{\rm s}$ structures (Fig.~\ref{geo}) is the tilt angle,
which is smaller in the PW91 relaxation. The two spectra are quite similar in
both shape and intensity: minor differences are observed only in the the first
peak, that is shifted upward by 0.15\,eV in the structure shown in the right
panel of Fig.~\ref{geo}.

The combined effects of stacking and H-bonding on the absorption spectra are
simultaneously present in Fig.~\ref{stack} (lower panel), that corresponds to
two stacked GC$_{\rm H}$ pairs with the G's on the same strand of the parent
polymer \cite{gao}.  The shape of the spectrum appears very similar to that of a
single GC$_{\rm H}$ pair (Fig.~\ref{hbond}, top left), but the intensity of all
peaks is reduced. This decrease is a typical qualitative effect of the stacking
arrangement, although the molecular details fix the quantitative aspects (we
described the stacking between guanine and cytosine above, whereas we are now
focusing on the stacking between two guanines).  In fact, in the low-energy
range we find again very slight shifts for the first two peaks, downward for
the first and upward for the second, as in Fig.~\ref{hbond} (top left). However,
the intensity of the second peak is equal to that obtained from the G+C
combination, whereas in the single GC$_{\rm H}$ pair it was higher. This is due
to a depression of the second peak (guanine) induced by stacking.

In the high-energy range, the hypochromicity due to both the $\pi$-stacking and
the H-bonding couplings is confirmed. For the main peak located at 6.25\,eV we
find an intensity reduction by 62\% with respect to the free monomers.  The two
kinds of base couplings seem to act separately and independently, in the sense
that one does not affect the other: such separation was already reported
concerning the electronic structure \cite{difelice02}.

Summarizing this section, we have shown that stacked and H-bonded GC pairs
present slight differences in the absorption spectra, both in the low-energy and
high-energy ranges.  Hypochromicity has been found in both configurations and
is largest for the stacked pairs. Hypochromicity is very useful
because this intensity change can be used to follow the melting of the secondary
structure of nucleic acids when varying the temperature or environmental
parameters.

\section{Summary and perspectives}
\label{conclusion}

In this work we presented a complete study of the optical absorption spectra of
the five isolated gas-phase nucleobases and their assemblies: simple
Watson-Crick pairs, simple $\pi$-stacks of two bases and more complex $\pi$-stacks
of Watson-Crick GC pairs. These calculations of isolated simple
nucleobase-assemblies are a remarkable playground to prove the reliability of
TDDFT for DNA-based materials. Remarkably, for the first time the optical 
properties were computed by an ab-initio method for a helical conformation of two 
stacked GC pairs, where H-bonding and $\pi$-stacking effects are active 
simultaneously and can be distinguished.

The results can be summarized as follows. For the isolated bases we get spectra
in good agreement with previous theoretical works and (qualitatively) with
experiments. We reproduce the proper ordering of the $\pi \pi^*$ excitations,
namely the excitation energy increases in going from C to G to A. Moreover,
the LUMO state has always a $\pi$-like character whereas the HOMO is $\pi$-like
for the purines and $\sigma$-like for the pyrimidines. As concerns the base
assemblies (Watson-Crick H-bond pairs and stacked configuration), we obtain that
the shape of the spectrum is not much altered by the $\pi$-stacking or H-bond
interactions. However, we always get hypochromicity in the high energy range of
the spectrum. The hypochromicity induced by $\pi$-stacking is larger than that 
induced by H-bonding. For light polarized perpendicular to the bases we
get a blue-shift of the spectra compared to the spectra of the isolated bases.
In the stacked case, the HOMO and LUMO states are distributed both on the purine
and pyrimidine bases, whereas in the H-bonded configuration the HOMO is in the
purine and the LUMO in the pyrimidine (charge transfer-type excitation). When
combining both H-bonding and $\pi$-stacking, the two effects add independently,
and the hypochromicity in the UV is enhanced.

At this stage it is relevant to note that all our calculations are in the gas
phase, which means that solvent and environment effects are not taken into
account.  This hinders a direct comparison with experimental data, as most of
the available results correspond to nucleobases assemblies in solution.
However, several experimental and theoretical studies (including simplified
models of the solvent) showed that the $\pi\pi^{*}$ transitions are only
slightly affected by the presence of the solvent
\cite{mennucci,mishra,shukla02}: precisely, they 
are insensitive to the polarity of the solvent, but may 
undergo red-shifts. 
On the contrary, the ``dark'' $n\pi^{*}$ transitions turn
out to be significantly sensible to the polarity of the solvent
\cite{mennucci,mishra,shukla00,fulscher97,shukla02,shukla02c}. As shown in
Figs.~\ref{basesa} and~\ref{basesb}, the absorption spectra of the DNA-bases is
extremely anisotropic, due to their quasi-planar structure.  Consequently,
nearly all the oscillator strength is concentrated on the $\pi\pi^{*}$
transitions (orders of magnitude more intense than the $n\pi^{*}$ transitions).
Those transitions are excited only for light polarized in the base plane.
Therefore, these  results can be used to discriminate (albeit
qualitatively) which features are mostly limited to the gas phase (those
polarized out-of-plane) and which ones can be considered intrinsic to the
DNA-complex \cite{note_reviewer65}. 
Furthermore, the role of charge-transfer states needs of be
analyzed more in detail in order to understand their impact in the excited state
dynamics of DNA-based complexes (note that those states are very likely to be
dark, or with very low oscillator strength for light-induced electronic
excitations).

Besides optical absorption, there is another
optical technique that is widely used for the characterization of chiral
biomolecules -- circular dichroism. The computational method used in the present work
allows for a straightforward calculation of the rotatory power or circular
dichroism spectra keeping the simplicity of the time-propagation scheme, i.e,
without the need for empty states (see Ref.~\onlinecite{yabana99} for the
details). The implementation and computation of circular dichroism spectra by TDDFT 
is in progress and will be the topic of a self-standing investigation \cite{dicro}, 
that should allow the identification of helical fingerprint in the optical characteristics, 
and more direct interpretation of standard post-synthesis experimental data.

\section*{Acknowledgements} 

The authors thank X. Lopez and E. Sansebastian for helpful discussion.
Funding was provided by the EC through grant IST-2001-38951, 
by the EC 6th Framework Network of Excellence NANOQUANTA 
(NMP4-CT-2004-500198), by Spanish MEC and
by the INFM through the Parallel Computing Committee.
A. Rubio acknowledges the Humboldt Foundation under the Bessel
research award (2005). We thank computer time provided by the 
SGI of the UPV and DIPC as well as the Barcelona supercomputer center.


\begin{thebibliography}{10}
\bibitem{note_reviewer65} With this reasoning, we do not claim that the values 
of the $\pi\pi^{*}$ transitions are to be considered exact also for the 
solution phase, but they are not expected to undergo major modifications in 
different solutions. Most  likely, the effect of the environment on the 
$\pi\pi^{*}$ transitions can be expected 
to be a systematic shift toward low energies, that should be taken into 
account  for a more strict comparison in Tables I-VI. However, such a shift 
cannot be computed at present at the same level of theory.
\end{thebibliography}
\end{document}